\documentclass[12pt]{iopart}
\expandafter\let\csname equation*\endcsname\relax
\expandafter\let\csname endequation*\endcsname\relax
\usepackage{amsmath,amssymb,bm,dsfont}
\usepackage{cite,txfonts,comment,braket}
\usepackage{graphicx,color}
\usepackage[colorlinks=true,linkcolor=blue,urlcolor=blue,citecolor=blue]{hyperref}

\begin{document}

\title{\boldmath Glassy disorder-induced effects in noisy dynamics of Bose-Hubbard and Fermi-Hubbard systems}

\author{Saubhik~Sarkar$^{1,2}$ and Ujjwal~Sen$^{2}$}

\address{Institute for Quantum Science and Technology and Department of Physics and Astronomy, University of Calgary, Calgary, Alberta T2N 1N4, Canada}
\address{Harish-Chandra Research Institute, A CI of Homi Bhabha National Institute, Chhatnag Road, Jhunsi, Allahabad 211 019, India}

\begin{abstract}
We address the effects of quenched disorder averaging in the time-evolution of systems of ultracold atoms in optical lattices in the presence of noise, imposed by an environment. For bosonic systems governed by the Bose-Hubbard Hamiltonian, we quantify the response of disorder in Hamiltonian parameters in terms of physical observables, including bipartite entanglement in the ground state, and report the existence of disorder-induced enhancement in weakly interacting cases. For systems of two-species fermions described by the Fermi-Hubbard Hamiltonian, we find similar results. In both cases, our dynamical calculations show no appreciable change in the effects of disorder from that of the initial state of the evolution. We explain our findings in terms of the statistics of the disorder in the parameters and the behaviour of the observables with the parameters.   
\end{abstract}

%\maketitle
\ioptwocol

% =============================================================================
\section{Introduction}
\label{Introduction}

Over the past few decades, development in the experimental realisation of quantum systems has led to a substantial enrichment of our understanding of strongly-correlated quantum many-body systems~\cite{Bloch2006,Bloch2008,Lewenstein2007,Lewenstein2017}. Quantum simulators have been successfully implemented in different physical systems including optical lattices~\cite{Cirac2012,Bloch2012}, photons~\cite{Guzik2012,Northup2014}, ion-traps~\cite{Leibfried2003,Blatt2012}, and superconducting qubits~\cite{Makhlin2001,Clarke2008}. This enables one to study several interesting physical observables in a many-body scenario, including magnetic order and entanglement~\cite{Horodecki2009}. Presence of disorder in the physical realisations, in general, suppresses such properties but in special cases, counter-intuitive effects like disorder-induced enhancements have also been observed~\cite{Villain1980,Henley1989}. This has been observed for physical quantities like magnetisation, and bipartite as well as multipartite entanglement in spin systems~\cite{Villa2014,Misguich2013,Minchau1985,Moreo1990,Feldman1998,Volovik2006,Abanin2007,Adamska2007,Santos2004, Monasterio2005,Lakshminarayan2005,Karthik2007,Brown2008,Dukesz2009,Hide2009,Prabhu2011,Sadhukhan2016,Mishra2016}, two-component Bose-Einstein condensates~\cite{Niederberger2008}, topological superconductors~\cite{Foster2014}, and Anderson-localised systems~\cite{Sandoval2006,Fujii2010}. It is also prevalent in systems governed by Hubbard Hamiltonian where the disorder can be implemented, for example, in the nearest neighbour tunnelling parameter, on-site interaction energy, or on-site chemical potential~\cite{Niederle2015}.

In the case of disordered site-specific chemical potential, study of entanglement in Hubbard systems is quite widespread. For Bose-Hubbard systems, ground state phase transitions as well as excited state phase transitions (many-body localization) can be probed using single-site entanglement~\cite{DellAnna2011}, entanglement spectra~\cite{Deng2013,Wahl2019}, bipartite entanglement~\cite{Andraschko2014,Goldsborough2015,Elben2018}, and entanglement area law~\cite{Yu2018}. Similar studies have also been conducted on Fermi-Hubbard systems~\cite{Franca2011,Karlsson2014,Smith2018}. On the other hand, disorder in local energy offset~\cite{Zhou2010} can also give rise to similar effects on the orders of the particular phases such as the superfluid~\cite{Gerster2016,Abreu2018} and the Mott insulator phase~\cite{Lacki2011}. The other interesting disorder-induced phenomena include existence of phases like Bose glass~\cite{Alvarez2103,Yao2014} (analogous to spin glass~\cite{Aharony1978,Tsomokos2011,Binder1986,Belitz2005,Das2008,Alloul2009}), where unitary glassy dynamics has also been studied~\cite{Caballero2013Glassy, Yan2017Dynamics}. As most of the studies so far generally have considered disorder in local energies, one natural direction to explore is the behaviour of the measures of order and two-site entanglement in the cases of disorder in the other Hamiltonian parameters which can also be implemented in physical set-ups~\cite{Niederle2015,Gopalakrishnan2011}. We consider disorders that stay unchanged during the dynamics of the system and hence, are of \textit{glassy} nature due to the long equilibration time. Note that this is not related to the aforementioned glass phases in Hubbard models~\cite{Fisher1989Boson,Giamarchi2001Competition}. 

Advances in the experimental techniques in recent times can be used to study in detail the effect of disorder in these systems in well-controlled conditions. In particular, the experiments with cold atoms in optical lattices have proved to be exceptionally effective tools that provide with an ability to realise these quantum systems and carry out precise measurements on them, due to the remarkable control over the system parameters. Efficient isolation from the environment enables observation of coherent dynamics as well as non-equilibrium dynamics~\cite{Cazalilla2011}. In laboratories, the successful observations of coherent as well as non-equilibrium dynamics depends on how well the isolation is between the system and the environment. Nevertheless, in any experimental set-up, noise cannot be fully avoided. For example, technical noise like the amplitude fluctuations of the optical lattice potential causes fluctuation in the Hamiltonian parameters. Temporal fluctuations can cause site-independent disorder whereas spatial fluctuations can cause site-dependent disorder. On the other hand, a major source of noise of quantum nature arises from spontaneous emission events due to coupling to the vacuum modes of the electromagnetic radiation. Therefore, it is important to study the evolution of the system in the presence of such noises.

In this work, we investigate the behaviour of quenched averaged physical properties of ground states of quenched disordered physical systems as well as their dynamical evolution when exposed to environmental noise. We specifically look at two different systems, viz.~of bosons and of two-species fermions, loaded in the lowest band of a one-dimensional optical lattice, described by single-band Bose-Hubbard~\cite{Jaksch1998} and the Fermi-Hubbard Hamiltonians~\cite{Esslinger2010}, respectively. We investigate the reaction to quenched disorder imposed individually in the tunnelling parameter and the on-site interaction on the ground state properties. The effect we seek to find is a disorder-induced enhancement in these quantities and we do find it for particular parameter ranges. We then go on to study the dynamics of these effects in presence of spontaneous emission events which we do by evolving the corresponding master equation for the system density operator~\cite{Lehmberg1970,Lehmberg1970a,Gordon1980,Dalibard1985, Gardiner2010,Pichler2010,Sarkar2014}. The disorder-induced effects are found to be sustained, at least in terms of short-time behaviours. For the numerical results on large systems, where exact diagonalisation is not possible, we use a combination of density matrix renormalisation group (DMRG) algorithm~\cite{Vidal2004,Daley2004,White2004,Verstraete2008,Schollwock2011} and the quantum trajectory method~\cite{Daley2014,Carmichael1993,Molmer1993,Dum1992} to compute the dynamics.   

The article is organised as follows. In section~\ref{Model Hamiltonians and observables}, we discuss the model Hamiltonians and lay out the physical observables we compute. In section~\ref{Quenched disorder and quenched averaging}, we briefly explain the concept of quenched disordered systems and the process of quenched averaging. We show our findings in the ground states of our model Hamiltonians in section~\ref{Effect of disorder in the ground state}. In section~\ref{Dynamics of quenched averaged observables}, we show the dynamical results in an open quantum system. In section~\ref{Site-dependent disorder}, we discuss the case where the disorder in tunneling parameter is site-dependent. We present our concluding remarks in section~\ref{Conclusion}.

% =============================================================================
\section{Model Hamiltonians and observables}
\label{Model Hamiltonians and observables}

The Bose-Hubbard Hamiltonian in 1D that describes bosons in the lowest band of an optical lattice can be represented as
\begin{align}
H_{BH} = - \displaystyle\sum_{i} \left(J_i b^{\dag}_{i} b_{i+1} + \text{H.c.}\right) + \displaystyle\sum_i \frac{U_i}{2} n_{i} \left(n_{i}-1\right),
\label{BH}
\end{align}
where $J_i$ is the tunnelling parameter between sites $i$ and $i+1$, $b_{i}$ ($b^{\dag}_{i}$) is the bosonic annihilation (creation) operators for the $i$-th site, $U_i$ is the on-site interaction strength and $n_{i}$ is the number operator $b^{\dag}_{i}b_{i}$ for the $i$-th site. On the other hand, the single-band two-species (with spins $\uparrow$ and $\downarrow$ ) Fermi-Hubbard Hamiltonian in 1D, we choose to work with, is
\begin{align}
H_{FH} = - \displaystyle\sum_{i,s} \left(J_i c^{\dag}_{i,s} c_{i+1,s} + \text{H.c.}\right) + \displaystyle\sum_i U_i n_{i,\uparrow} n_{i,\downarrow},
\label{FH}
\end{align}
where $J_i$ is again the tunnelling parameter between sites $i$ and $i+1$ and $U_i$ is the on-site interaction energy for two fermions with different spins at site $i$. The annihilation (creation) operators $c_{i,s}$ ($c^{\dag}_{i,s}$) for site $i$ and spin $s$ obey the fermionic anti-commutation relations. The number operator in this case is $n_{i,s}=c^{\dag}_{i,s}c_{i,s}$, and for each spin species, it has eigenvalues $0$ and $1$, due to Pauli exclusion principle. The site index in the subscript of $J_i$ and $U_i$ are dropped when these parameters are site-independent and only change from one realization to the other.

In our calculation, we impose quenched disorder on both the Hamiltonian parameters, $J_i$ and $U_i$, and examine the cases where the disorder is only in the interaction term or only in the tunnelling term. In each of these cases, the physical quantities we compute, are described in the following. For the Bose-Hubbard Hamiltonian, a quantum phase transition from the superfluid to the Mott insulator phase~\cite{Fisher1989} occurs at $U/J \approx 3.37$ for unit filling in the 1D case~\cite{Kuhner2000}. This phase transition can be detected by looking at the trends of the single-particle density matrix (spdm), $\langle b^{\dag}_{i}b_{j} \rangle$. In the superfluid phase, spdm elements show polynomial decay with distance between the sites, $|i-j|$, whereas in the Mott insulator phase, the spdm elements fall off exponentially with distance~\cite{Daley2005}. For our work, we define two Mott orders, averaged over the lattice sites, given by   
\begin{align}
M_1=\frac{1}{N} \displaystyle\sum_{i=1}^N \frac{\left|\left\langle b^{\dag}_{i}b_{i}\right\rangle \right|} {\left|\left\langle b^{\dag}_{i}b_{i+1}  \right\rangle \right|},
\end{align}
and
\begin{align}
M_2=\frac{1}{N} \displaystyle\sum_{i=1}^N \frac{\left|\left\langle b^{\dag}_{i}b_{i}\right\rangle \right|} {\left|\left\langle b^{\dag}_{i}b_{i+2}  \right\rangle \right|}. 
\end{align}
One can check how these Mott orders scale with interaction strength in the ordered case. We observe that in the superfluid phase, $M_1 \sim (U/J)^{1.5}$ and $M_2 \sim (U/J)^{1.4}$. In the Mott insulator phase they both scale quadratically with $U/J$.

For the fermions, since the two-species Fermi-Hubbard Hamiltonian is a paradigmatic system to study antiferromagnetic order for repulsive interaction~\cite{Cirac2012,Bloch2012}, we consider two measures of spin correlation functions that quantify magnetic order. Defining the $z$ component of spin at the $i$-th site as $S_i^z=(n_{i,\uparrow} - n_{i,\downarrow})/2$, we write down the on-site spin correlation function, averaged over all the lattice sites, as
\begin{align}
S_0=\frac{1}{N} \displaystyle\sum_{i=1}^N \left\langle \left(S_i^{z}\right)^2 \right\rangle.
\label{S_0} \end{align}
The other quantity which we study is the spin correlation between two adjacent sites. When using open boundary conditions we calculate this in the middle of the chain, given by
\begin{align}
S_1 = \left\langle S_{\frac{N}{2}}^z S_{\frac{N}{2}+1}^z \right\rangle,
\end{align}
and only small quantitative changes can be found in the result when the averaging is done over the whole chain. As a particularly fitted measure of the bipartite entanglement, we calculate the logarithmic negativity, $\mathcal{LN}$~\cite{Karol1998,Jinhyoung2000,Vidal2002,Plenio2005}. The negativity, $\mathcal{N}$, of a bipartite system $\rho_{AB}$, comprising of sub-systems $A$ and $B$, is defined as the absolute value of the sum of negative eigenvalues of $\rho_{AB}^{T_A}$, the partial transpose of $\rho_{AB}$ with respect to $A$~\cite{Peres1996,Horodecki1996}. In terms of the trace norm of $\rho_{AB}$, defined as $|| \rho_{AB}^{T_A} ||_1 = \text{Tr} \sqrt{\rho_{AB}^{T_A \dag} \rho_{AB}^{T_A}}$, negativity can be rewritten as
\begin{align}
\mathcal{N}=\frac{|| \rho_{AB}^{T_A} ||_1 -1}{2},
\end{align}    
and logarithmic negativity is then defined as
\begin{align}
\mathcal{LN}=\text{log}_2 \left( 2 \mathcal{N}+1\right).
\end{align}
We evaluate these quantities to examine the response of the quenched disorder averaging process which we describe next.

% =============================================================================
\section{Quenched disorder and quenched averaging}
\label{Quenched disorder and quenched averaging}

\begin{figure}[t]
\centering
\includegraphics[width=0.6\linewidth]{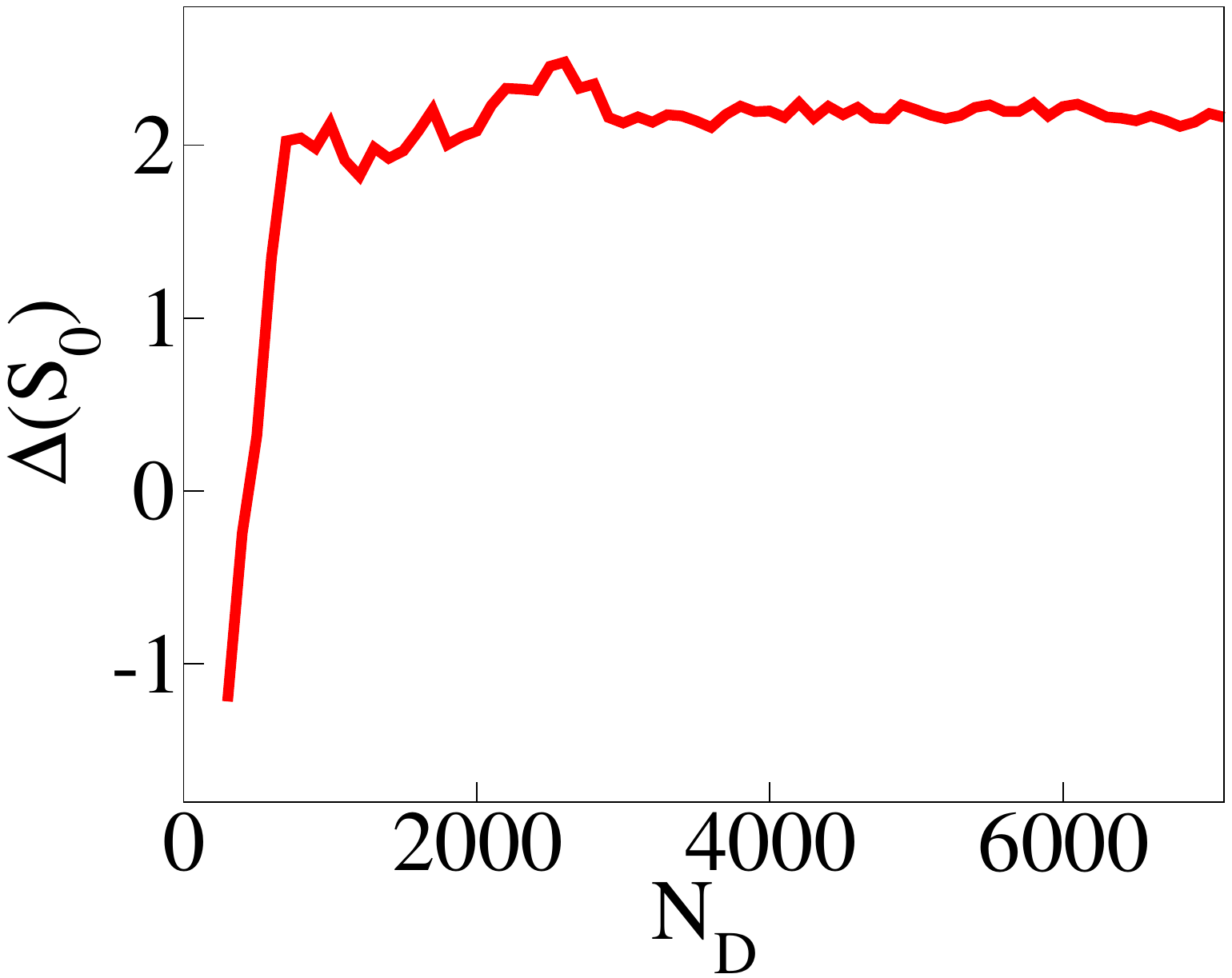}
\caption{Convergence plot for $\Delta(S_0)$ with the number of random disorder realisations, $N_D$. This plot shows the result for the ground states of the Fermi-Hubbard Hamiltonian in equation~\eqref{FH} with $U=2J_0$ for a system size of $N=32$ computed using DMRG method with bond dimension $256$, and for the case where the disorder is only in the tunnelling term with average value $J_0$ along with $\sigma=0.5J_0$ and $\mu=0$ (equation~\eqref{J_disorder}).}
\label{conv}
\end{figure}

In an ordered system, the Hamiltonian parameters have constant values throughout the time during which it is observed, and do not change from one realisation to another. This does not hold for the disordered case. An important class of disordered systems are those in which the Hamiltonian parameters are `quenched', in the sense that the typical time-scales required for the parameters to equilibrate are much larger than the time-scales of the system dynamics that is experimentally considered. Systems with quenched disordered parameters are often referred to as \textit{glassy}~\cite{Aharony1978,Villain1980,Minchau1985,Henley1989,Prabhu2011,Sadhukhan2016,Mishra2016}. Therefore, for all relevant purposes, the value of a particular disordered parameter remains constant during the time considered for observation of the dynamics. However, the next (or any other) realisation of the same system has a value of the same parameter that is independent from but identically distributed as that parameter in the previous case. For example, if we want the disorder only to be in the tunnelling term we take
\begin{align}
J = J_0 + \delta J(\mu,\sigma),
\label{J_disorder}\end{align}   
where in this work we take $\delta J(\mu,\sigma)$ from a set of independent and identically distributed Gaussian random variables with mean $\mu$ and standard deviation $\sigma$. We consider different $\sigma$-s in order to investigate the effect of changing the width of the distribution, while $\mu$ is set to be zero. The results are found to be qualitatively similar and the effects get more pronounced with increase in $\sigma$. With this observation, we fix $\sigma=0.5J_0$, and use this value for all our results in this paper. For Gaussian random variables, the values of $\delta J(\mu,\sigma)$ typically fall between $\mu-3\sigma$ and $\mu+3\sigma$. For the tunnelling term we discard the negative values, which is around $1.25\%$ of all realisations. We first determine the ground state for a given set of values of $U/J_0$. For bosons, the calculations are performed at unit filling, and for fermions, the same is done for half-filling for each of the spin-species. For small systems ($N=8$), we use exact diagonalisation to find the ground state while the imaginary time evolution, using time evolving block decimation (TEBD) algorithm~\cite{Vidal2004} under the framework of DMRG methods, is used to find the ground state with larger number of lattice sites ($N=32$). The exact diagonalisation method is used with periodic boundary conditions which only means that there is a tunnelling terms connecting the two ends, and therefore does not contradict the presence of disorder. The quenched disorder averaging is done by finding the value of an observable, $\mathcal{O}$, for a given realisations, and then averaging over a large number of realisations. The effect of the disorder is quantified by percentage value of the relative difference of the averaged value with respect to its value in the ordered case, which we denote by $\Delta(\mathcal{O})$. Therefore,
\begin{align}
\Delta(\mathcal{O}) = \frac{\mathcal{O}_{\text{average}} - \mathcal{O}_{\text{ordered}}}{\mathcal{O}_{\text{ordered}}} \times 100.
\end{align}
The number of realisations, $N_D$, required to obtain the quenched average observables is also analysed, and the resulting convergence plot is shown in figure~\ref{conv}. The typical value of $N_D$ that needs to be considered to attain convergence up to third decimal place is found to be a few thousands. This plot shows the result obtained for $\Delta(S_0)$ (see equation~\eqref{S_0}) for the Fermi-Hubbard model with a system size $N=32$ computed using DMRG with bond dimension of $256$. In the succeeding sections, we report our findings for $5000$ realisations of disorder realisations.  

% =============================================================================
\section{Effect of disorder in the ground state}
\label{Effect of disorder in the ground state}

To investigate the effect of the quenched disordered averaging, we engage in the two scenarios in context of the two different terms, viz.~tunnelling and on-site interaction strengths, in the Hamiltonians. When the disorder is only present in the interaction term of the Hamiltonian in equation~\eqref{BH} and equation~\eqref{FH}, we do not observe any noticeable disorder-induced order. In particular, the quenched averaged values of observables for Gaussian distribution of the disorder realisations also appear to be Gaussian in nature and close enough to the ordered value of the observable. We therefore have an absence of disorder-induced enhancement. Hence, in further discussions, we only investigate the second scenario with disorder in the tunnelling term.

% =============================================================================
\subsection{Bose-Hubbard Hamiltonian}

\begin{figure}[t]
\centering
  \begin{tabular}{cc}
    \includegraphics[width=0.5\linewidth]{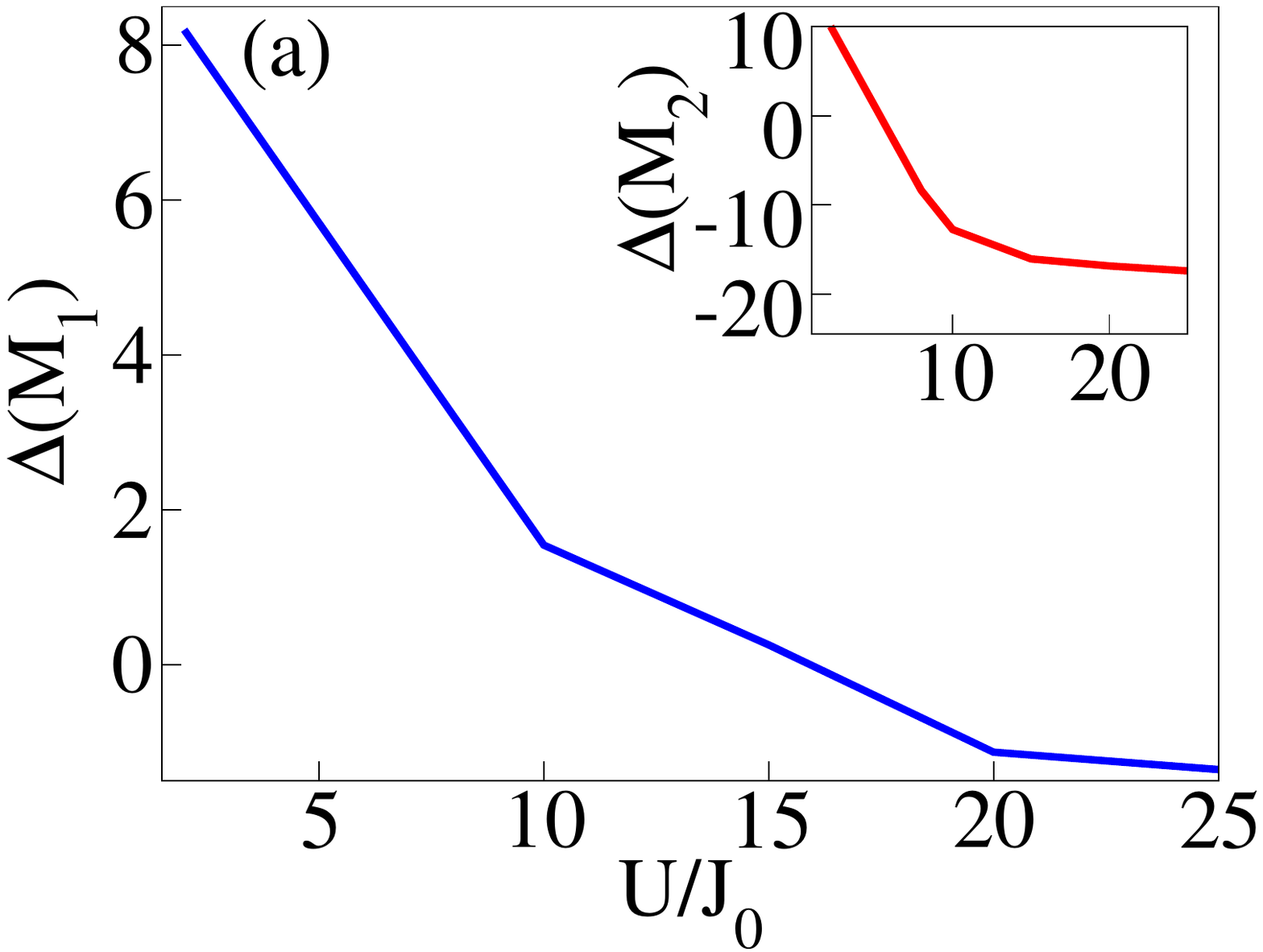}&
    \includegraphics[width=0.5\linewidth]{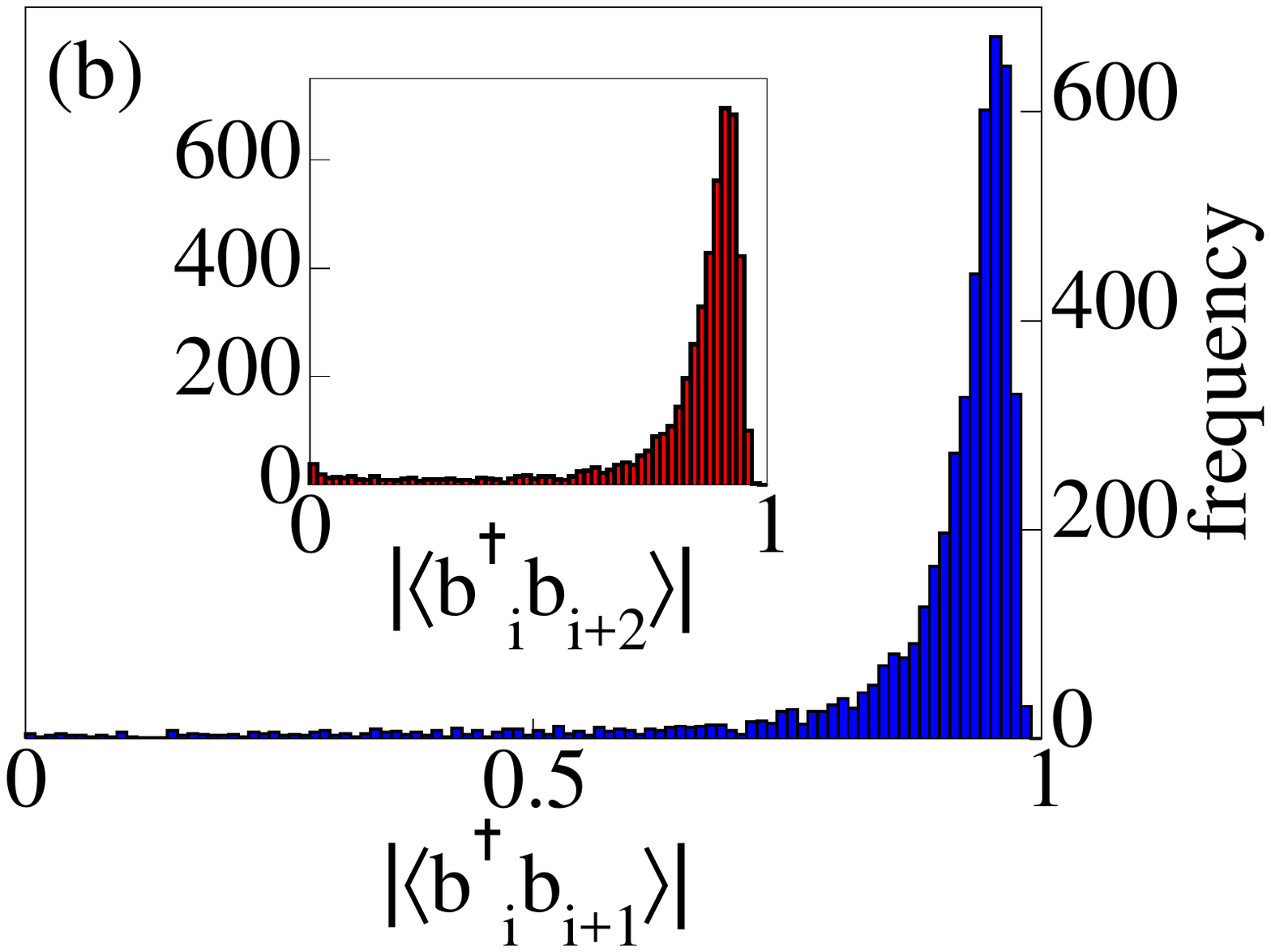}\\
    \includegraphics[width=0.5\linewidth]{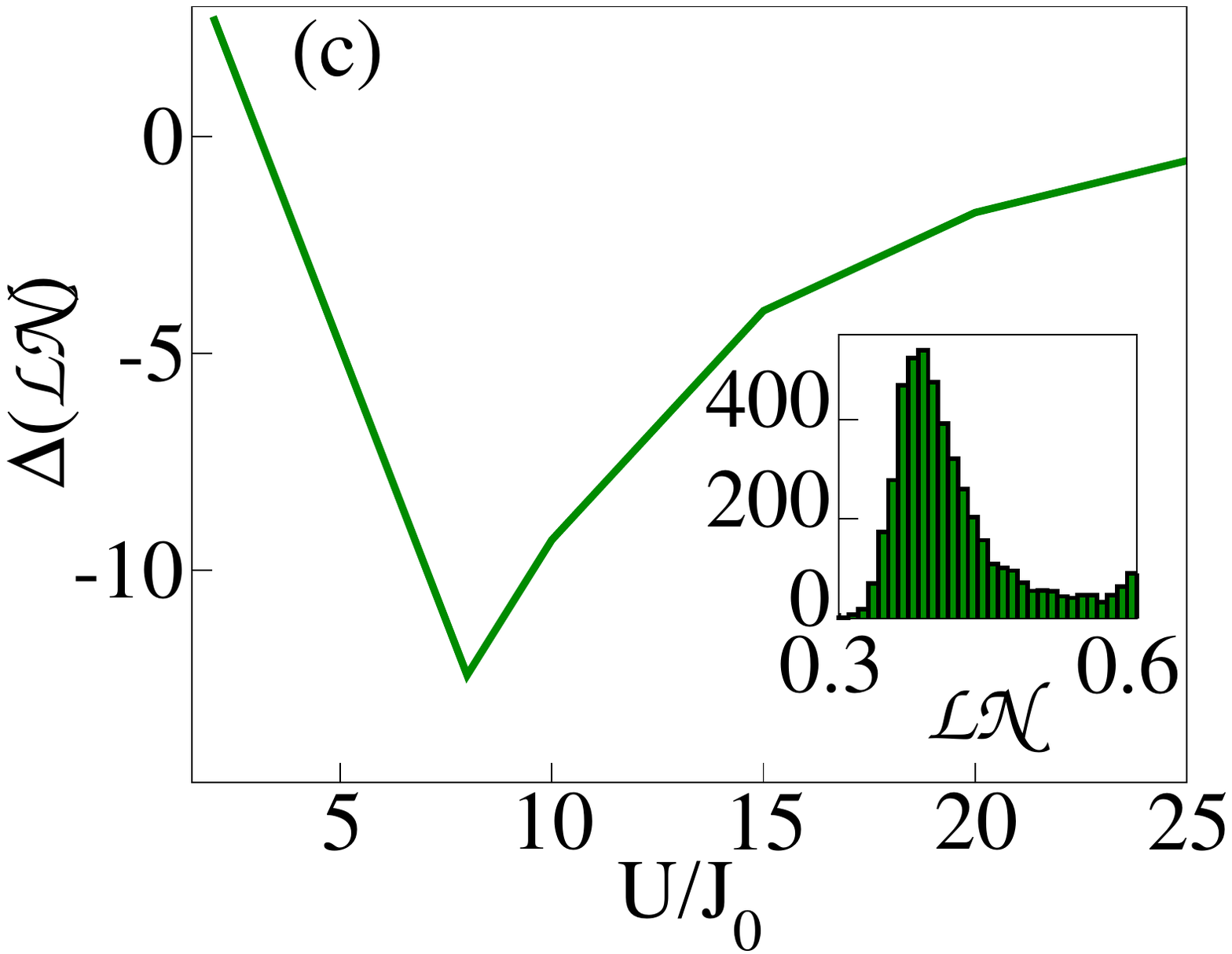}&
    \includegraphics[width=0.5\linewidth]{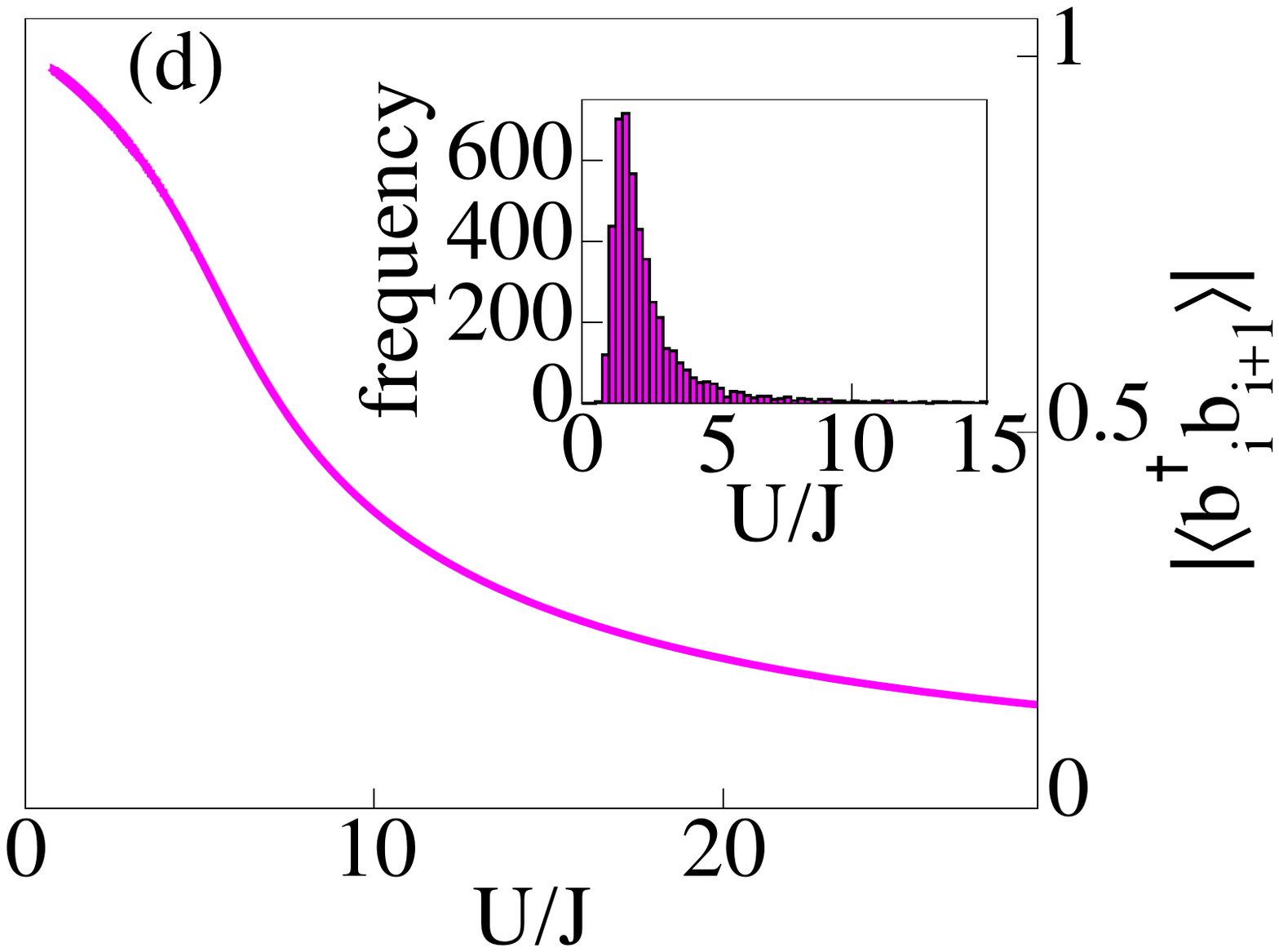}\\
    \includegraphics[width=0.5\linewidth]{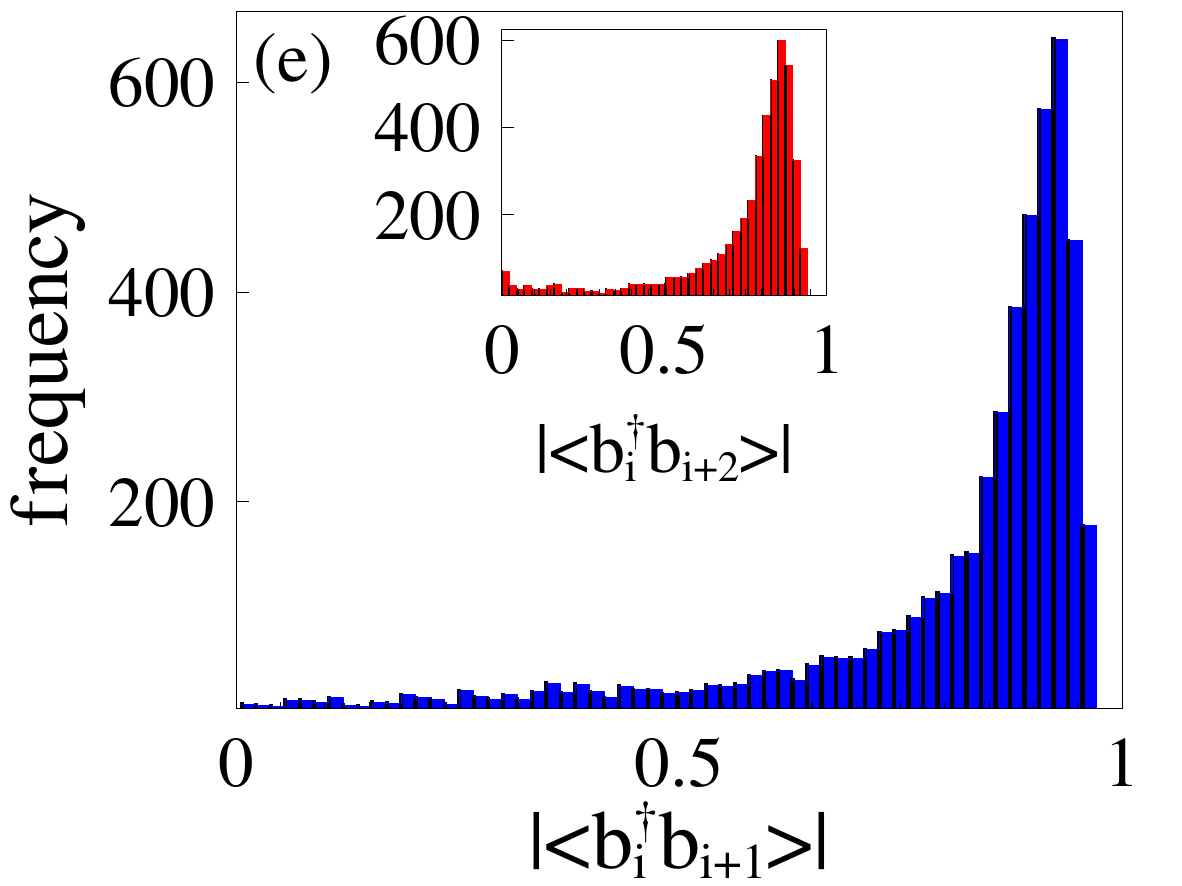}&
    \includegraphics[width=0.5\linewidth]{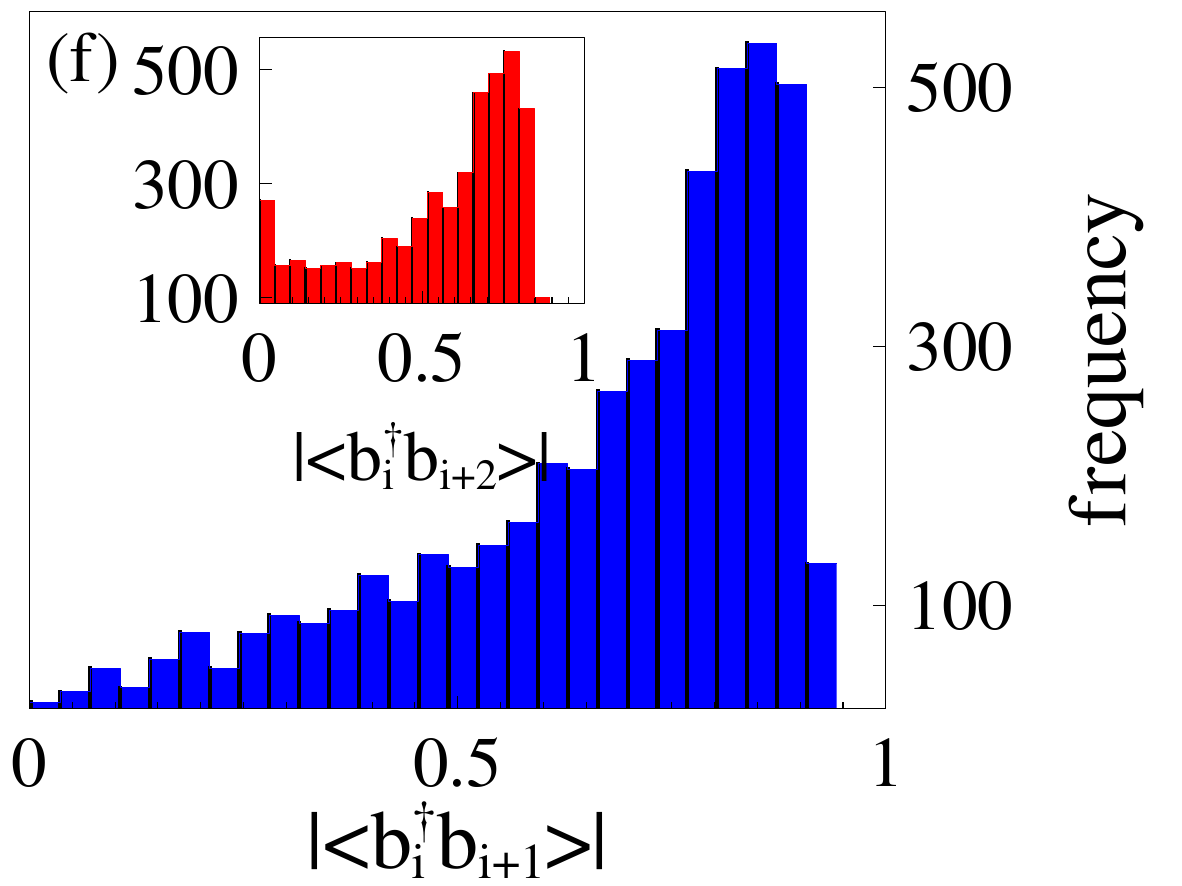}
  \end{tabular}
\caption{Effect of quenched disorder averaging in Bose-Hubbard model with the disorder realised only in the tunnelling term (equation~\eqref{J_disorder}). For plots (b), (c), and (d), the system size is $N=8$, where we have used exact digonalisation with periodic boundary condition. (a) $\Delta (M_1)$ as a function of $U/J_0$. The system size here is $N=32$ and the bond dimension used for the DMRG calculation is $256$. The inset shows similar plot for $M_2$. (b) Statistics for the $\langle b^{\dag}_{i}b_{i+1} \rangle$ that is required to compute the $M_1$ values for different disorder realisations with $U/J_0=2$. These statistics are $i$-independent as we use periodic boundary condition. The inset shows similar plot for $\langle b^{\dag}_{i}b_{i+2} \rangle$. Additional plots for such distributions are presented in (e) for $U/J_0=3$ and in (f) for $U/J_0=5$. (c) $\Delta (\mathcal{LN})$ as a function of $U/J_0$. The inset shows the statistics for the $\mathcal{LN}$ values for different disorder realisations with $U/J_0=2$. (d) Behaviour of $|\langle b^{\dag}_{i}b_{i+1} \rangle|$ as a function of $U/J$. The inset shows the statistical distribution of $U/J$ corresponding to the random Gaussian distribution of $J$ with average value $J_0$ and $\sigma=0.5J_0$. Here we have taken $U/J_0=2$.}
\label{BH_gs}
\end{figure}

When the disorder is in the tunnelling term of the Bose-Hubbard Hamiltonian, we detect presence of enhancement in suitable parameter ranges. For the ordered state, with $J=J_0$, and for the disordered states we first measure the Mott orders and the two-site entanglement in the ground state. We find disorder-induced enhancement (i.e.,~positive $\Delta$ values) for small enough values of $U/J_0$ which becomes smaller as we increase $U/J_0$ and eventually $\Delta$ becomes negative. This result is shown in figure~\ref{BH_gs}(a) where $\Delta (M_1)$, the Mott order in terms of adjacent sites for a $32$-site system, is displayed with $U/J_0$. We have taken $N_D=5000$ to obtain good convergence. The error bars are omitted as they tend to be quite small. Similarly, we find the Mott order in terms of next-neighbour sites, the findings of which are shown in the inset of figure~\ref{BH_gs}(a) where we plot $\Delta (M_2)$ with $U/J_0$. We also look into the behaviour of the logarithmic negativity $\mathcal{LN}$, as a measure of entanglement, taking two adjacent sites. The disorder effect can be found in figure~\ref{BH_gs}(c) where $\Delta (\mathcal{LN})$ is plotted as a function of $U/J_0$. Comparing figures~\ref{BH_gs}(a) and \ref{BH_gs}(c), we note that for $\Delta (M_1)$, the value of $U/J_0$ where the disorder-induced enhancement cease to exist, is larger than those in the cases of $\Delta (M_2)$ and $\Delta (\mathcal{LN})$. Therefore the enhancement is more pronounced in the Mott order of nearest-neighbour sites. Moreover, we notice that the superfluid phase shows more enhancement than the Mott insulator phase. Note that, these phases are in reference to the ordered parameter value $U/J_0$, around which the disorder is being considered. 

Towards understanding such observations, we analyze the statistics of the distributions of the different observables for the quenched disorder realisations. The distributions are displayed in figures~\ref{BH_gs}(b), (e), (f) and their insets, and also in the inset of figure~\ref{BH_gs}(c). We look at the distributions for $|\langle b^{\dag}_{i}b_{i+1} \rangle|$ (in connection to $M_1$), $|\langle b^{\dag}_{i}b_{i+2} \rangle|$ (in connection to $M_2$), and $\Delta (\mathcal{LN})$.  Although we mostly choose $U/J_0=2$ for plotting the distributions, the changes in the shapes of these asymmetric distributions are shown in figures~\ref{BH_gs}(b), (e), and (f) for $U/J_0 = 2, 3,$ and $5$, respectively, to display the behaviour across the phase transition. For a case where the value of $\Delta$ for an observable is positive, the average of its distribution (generated by the disorder) is greater than the corresponding value in the ordered case. Let us focus on a specific example. In order to explain the effect of disorder in the tunnelling term on the physical observable, $M_1$, it is instructive to look at the behaviour of $|\langle b^{\dag}_{i}b_{i+1} \rangle|$ as a function of $U/J$, in the \emph{ordered} case, shown in figure~\ref{BH_gs}(d). For small values of $U/J$ (e.g.~$U/J=2$), in the superfluid regime, we observe that the quantity $|\langle b^{\dag}_{i}b_{i+1} \rangle|$ falls faster than in the cases of higher values of $U/J$. This is because, the superfluid wavefunction gets affected rapidly by the localizing effects of the onsite interaction. The distribution of $U/J$, shown in the inset of figure~\ref{BH_gs}(d), is also an asymmetric one, where $U/J_0=2$, and \(J\) is Gaussian distributed with mean \(J_0\) and standard deviation \(J_0/2\). We now observe this distribution, along with the rate of change in the value of $|\langle b^{\dag}_{i}b_{i+1} \rangle|$  with $U/J$. We infer that in the disordered case, more weightage would appear in the disorder average from the smaller side of $|\langle b^{\dag}_{i}b_{i+1} \rangle|$ in figure~\ref{BH_gs}(d). Due to the fact that $|\langle b^{\dag}_{i}b_{i} \rangle|=1$ (uniform number density for periodic boundary condition), the quenched disordered average of $M_1$, therefore, is larger than the ordered value. On the other hand, as one goes deeper in the Mott insulator regime, where the ground state wavefunction is already localized, increasing the interaction does not have a strong effect on the spdm elements. Therefore, $|\langle b^{\dag}_{i}b_{i+1} \rangle|$ falls at a much slower rate with increase in $U/J_0$ which results in less weightage from the larger side of $|\langle b^{\dag}_{i}b_{i+1} \rangle|$. Hence, in that regime,  the quenched disordered average of $M_1$ is smaller than the ordered value, thereby showing no disorder-induced enhancement in this phase. This therefore provides the origin of the disorder-induced enhancement of \(M_1\) near the superfluid regime and its absence near the Mott one. Similar explanations hold for the disorder-induced enhancements and their absence for other observables as well.

% =============================================================================
\subsection{Fermi-Hubbard Hamiltonian}
 
\begin{figure}[t]
\centering
  \begin{tabular}{cc}
    \includegraphics[width=0.5\linewidth]{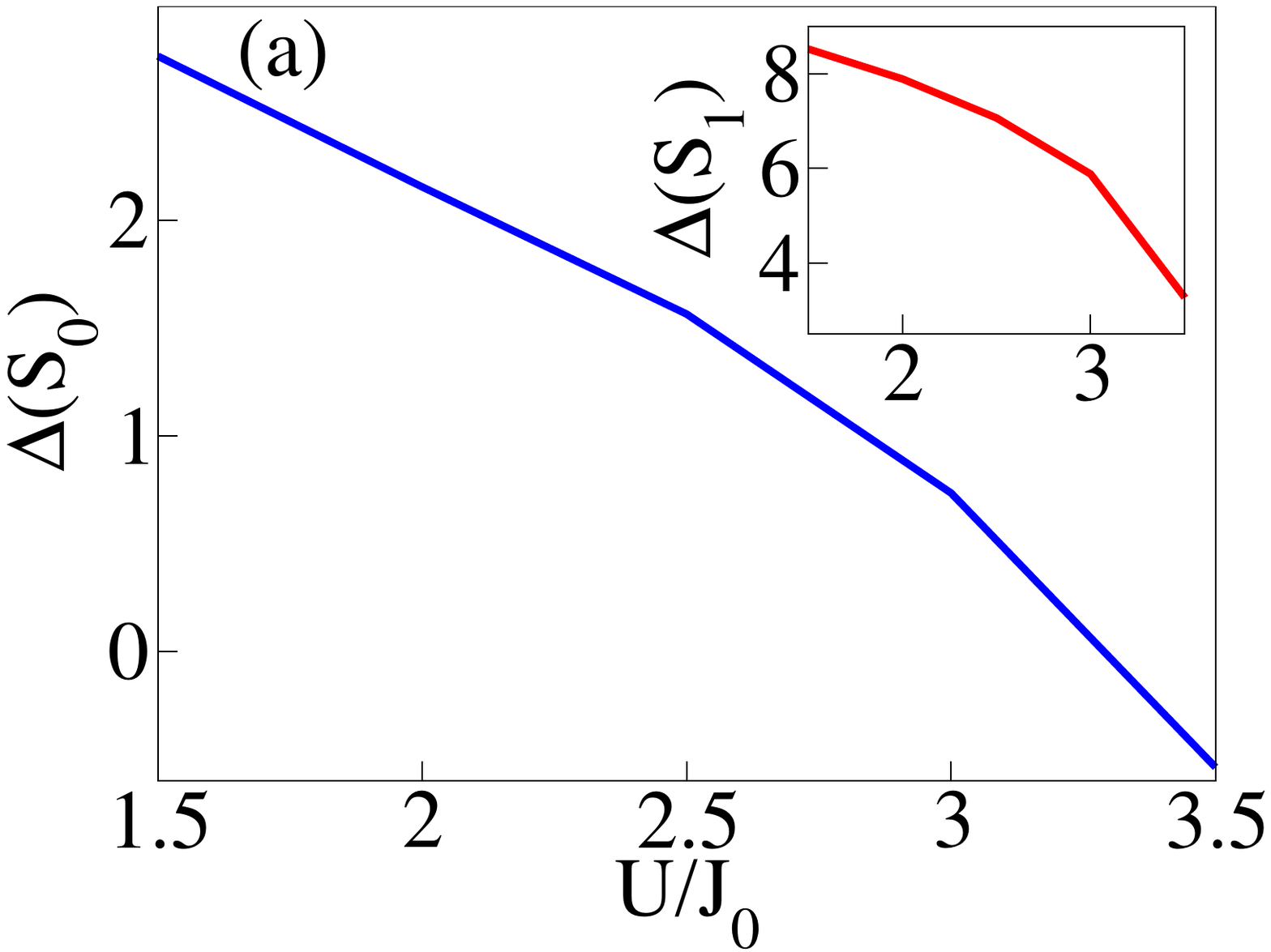}&
    \includegraphics[width=0.5\linewidth]{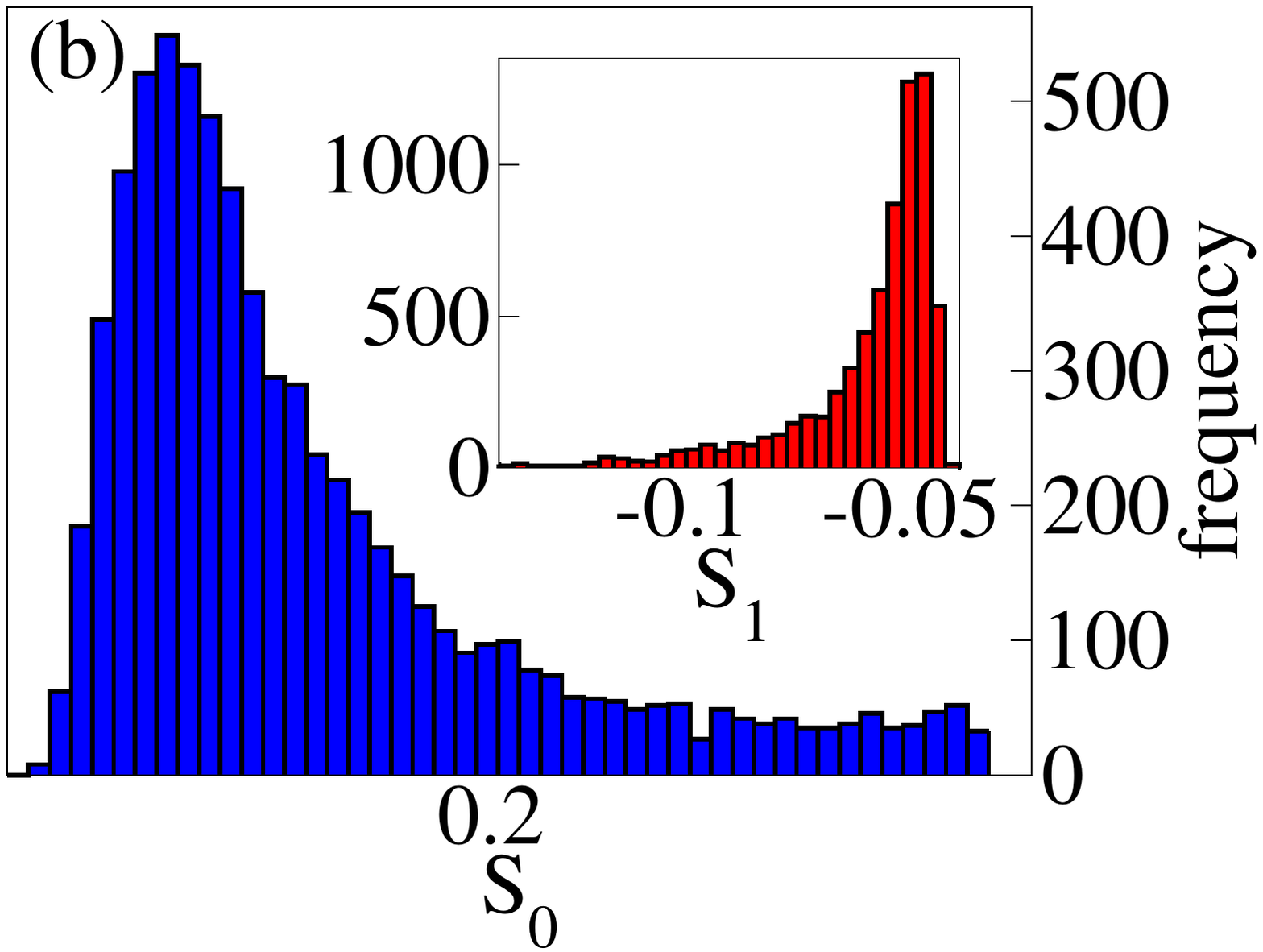}\\
    \includegraphics[width=0.5\linewidth]{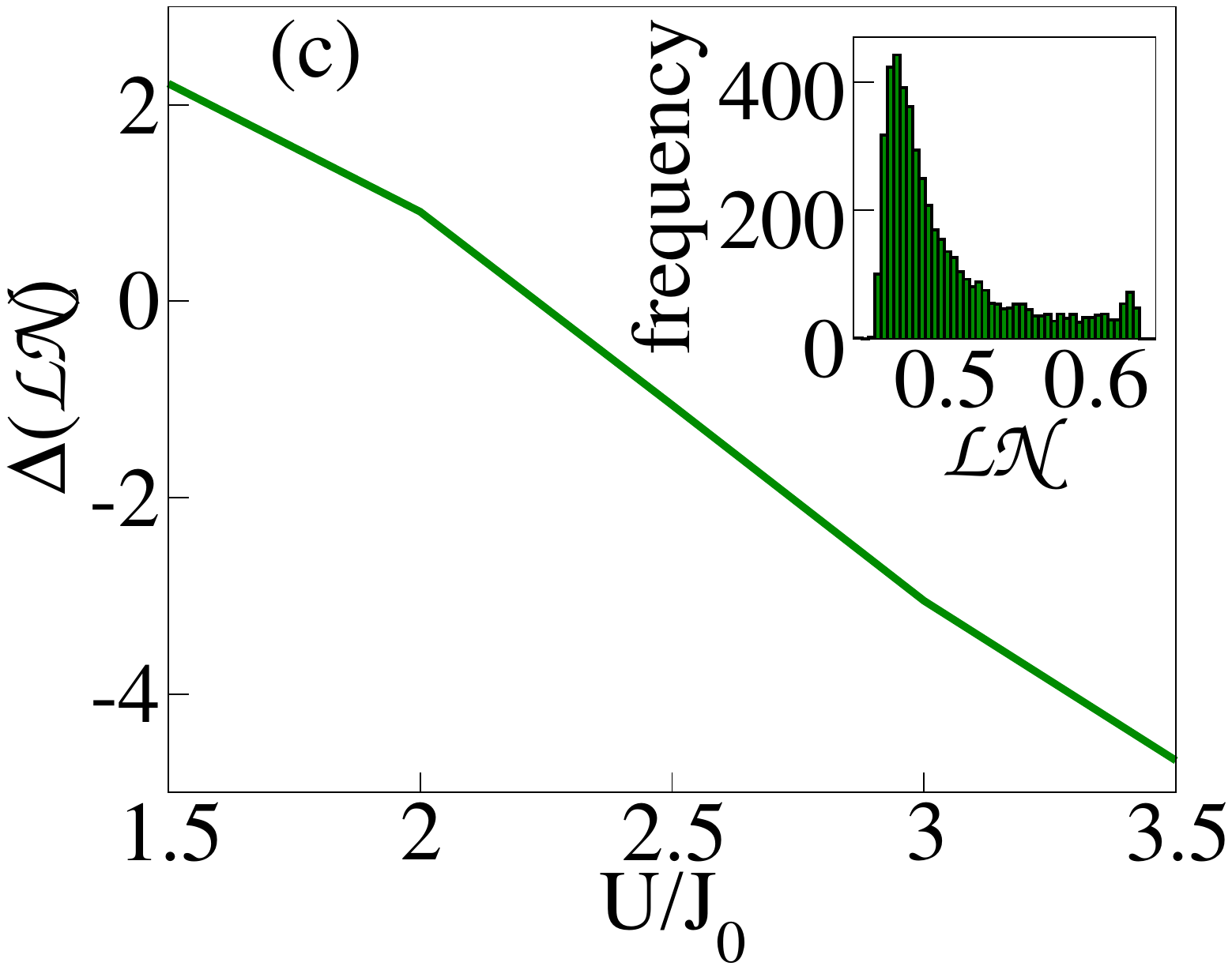}&
    \includegraphics[width=0.5\linewidth]{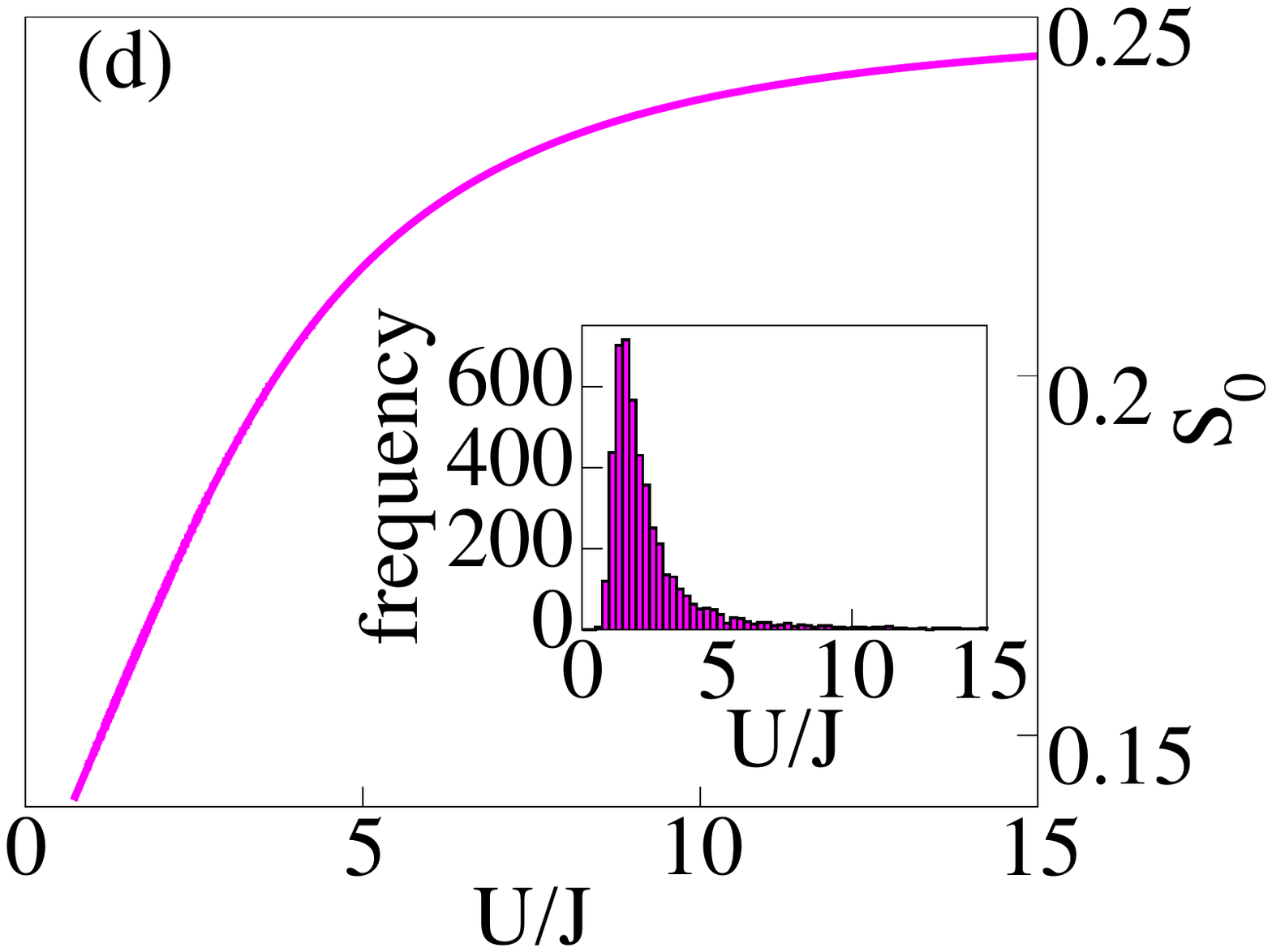}
  \end{tabular}
\caption{Effect of quenched disorder averaging in Fermi-Hubbard model with the disorder realised only in the tunnelling term (equation~\eqref{J_disorder}). (a) $\Delta (S_0)$ as a function of $U/J_0$. The system size is $N=32$ and the bond dimension used for the DMRG calculation is $256$. The inset shows similar plot for $S_1$. (b) Statistics for the $S_0$ values for different disorder realisations with $U/J_0=2$. The inset shows similar plot for $S_1$. (c) Disorder-induced order in Fermi-Hubbard model: $\Delta (\mathcal{LN})$ as a function of $U/J_0$. The system size is $N=8$. Statistics for the $\mathcal{LN}$ values for different disorder realisations with $U/J_0=2$. (d) Behaviour of $S_0$ as a function of $U/J$. The inset shows the statistical distribution of $U/J$ corresponding to the random Gaussian distribution of $J$ with average value $J_0$ and $\sigma=0.5J_0$. Here we have taken $U/J_0=2$.}
\label{FH_gs}
\end{figure}

We now study the patterns of correlation functions and bipartite entanglement in the Fermi-Hubbard Hamiltonian with a disordered tunnelling parameter. Like the Bose-Hubbard Hamiltonian, we find that for small values of $U/J_0$, the observables $\Delta (S_0)$, $\Delta (S_1)$, and $\Delta (\mathcal{LN})$ possess positive values, thereby showing enhancement of their values in presence of disorder (see figure~\ref{FH_gs}(a) and \ref{FH_gs}(c)). The investigation for the $32$-site system has been carried out by using DMRG techniques with bond dimension $256$. Again we perform the convergence check and conclude that $N_D=5000$ is sufficient for the quenched averaging. Both the antiferromagnetic order $S_1$ and logarithmic negativity $\mathcal{LN}$ have been calculated by considering the two adjacent sites in the middle of the chain. Similar to the Bose-Hubbard model, $U/J_0$ for which $\Delta (S_0)$ crosses zero is higher than that in the case of $\Delta (\mathcal{LN})$.

As before, we report the asymmetric statistical distributions of $S_0$, $S_1$, and $\mathcal{LN}$ with $U/J$ in figures~\ref{FH_gs}(b) and insets of \ref{FH_gs}(b) and \ref{FH_gs}(c) respectively. For these distributions, $U/J_0=2$, as before. The presence of disorder-induced enhancement for small $U/J_0$ values and its absence for large $U/J_0$ values can again be explained as before, in terms of the behaviour of the physical observables with $U/J$ and the distribution of $U/J$ itself (see figure~\ref{FH_gs}(d) and its inset). The rate of increase of $S_0$ is much larger for small $U/J_0$ values and therefore, due to the asymmetric distribution of $U/J$ values, more weightage is obtained from the larger side of $S_0$. This causes the quenched disordered average of $S_0$ to be bigger than the ordered value. Similarly, for larger $U/J_0$ values, the smaller rate of increase of $S_0$ causes the quenched disordered average of $S_0$ to be smaller than the ordered value. The different rates of change in the spin correlations with interaction strength can be understood from the fact that at strongly interacting regime, the ground state in the ordered case has anti-ferromagnetic order. The tunnelling term merely act as a perturbation and therefore the spin correlation changes slowly over a range of $U/J_0$. The correlations only start changing rapidly when the interaction and tunnelling terms become comparable, i.e., in the weakly interacting regime. 

For both bosons and fermions, although our studies are based on numerical simulations with moderate system sizes, the results obtained here give an intuitive understanding of observation of disorder-induced enhancements of observables in disordered systems. In particular, our calculation shows the presence of disorder-induced enhancements for the weakly interacting cases with quenched disorder in the tunnelling parameter. The enhancement can be observed in the bipartite entanglement and in the Mott orders for bosons and spin-correlations for fermions. 

% =============================================================================
\section{Dynamics of quenched averaged observables}
\label{Dynamics of quenched averaged observables}

After looking at the ground state properties, the next step is to evolve the system in time in the presence of noise that a real experimental set-up would suffer from. In the context of realising the Bose-Hubbard or Fermi-Hubbard Hamiltonian with cold atoms in optical lattices, an often unavoidable noise source is the spontaneous emissions, due to the atoms coupling to the vacuum modes of the electromagnetic field. This practically reduces a coherent atomic wavefunction to an incoherent mixture of localised Wannier states at different lattice sites, causing dephasing. Microscopic understanding for the mechanism of spontaneous emission enables one to write down the corresponding master equation for the atomic density operator under the suitable approximations, viz.~the dipole, Born-Markov, and rotating wave approximations. The resulting decoherence operators can be shown to be the number operators at each site both in the bosonic case~\cite{Pichler2010} and in the case two-species fermions~\cite{Greif2013,Cazalilla2009,Gorshkov2010,Scazza2014,Zhang2014,Cazalilla2014}. Therefore, the Lindblad form of the master equation for the system density operator $\rho$ in both cases can be represented as
\begin{align}
\dot{\rho}=-\frac{i}{\hbar}[H_{S},\rho]+\frac{\gamma}{2}\displaystyle\sum_{i}(2n_{i}\rho n_{i}-n_{i}n_{i}\rho-\rho n_{i}n_{i}),
\label{master}
\end{align}
where $H_S$ is the system Hamiltonian (i.e., $H_{BH}$ or $H_{FH}$), $\gamma$ is the effective decoherence rate, and the Lindblad operator $n_i$ is total number operator at site $i$. For two-species fermions, $n_i = n_{i,\uparrow} + n_{i,\downarrow}$. For each realization of the disordered Hamiltonian, we determine the ground state and then evolve it with equation~\eqref{master}. Due to exponential growth of Hilbert space dimension  number of sites, we compute the evolution of the density operator by the quantum trajectory method, also referred to as the Monte Carlo wavefunction method, that incorporates performing a stochastic average~\cite{Carmichael1993,Molmer1993}. In our work, exact diagonalisation for $N{=}8$ is combined with this algorithm. For $N{=}32$, we use a combination of DMRG algorithm with this method where individual trajectories are obtained by evolving the wavefunction under the effective Hamiltonian using the TEBD method~\cite{Daley2014}.

\begin{figure}[t]
\centering
  \begin{tabular}{cc}
    \includegraphics[width=0.5\linewidth]{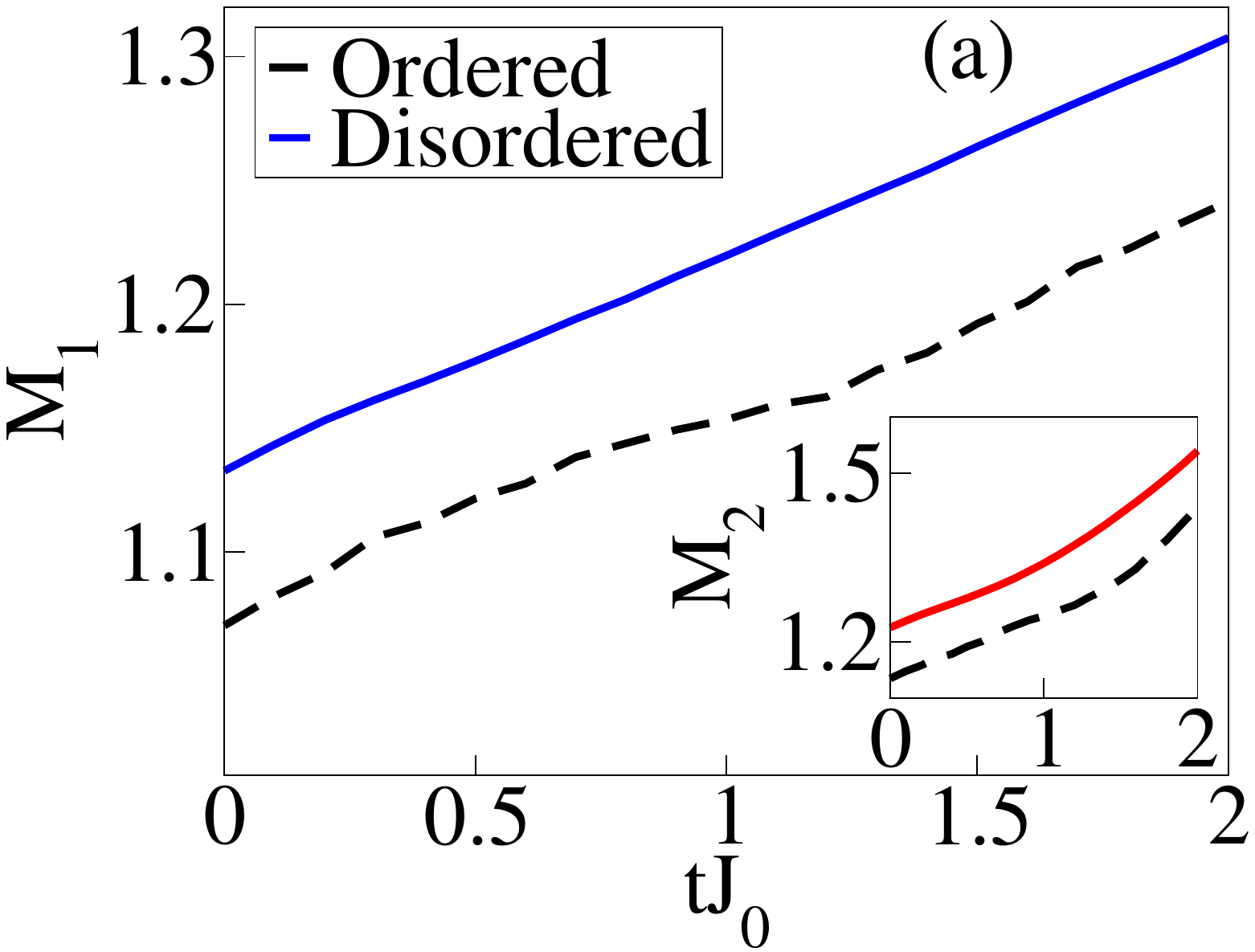}&
    \includegraphics[width=0.5\linewidth]{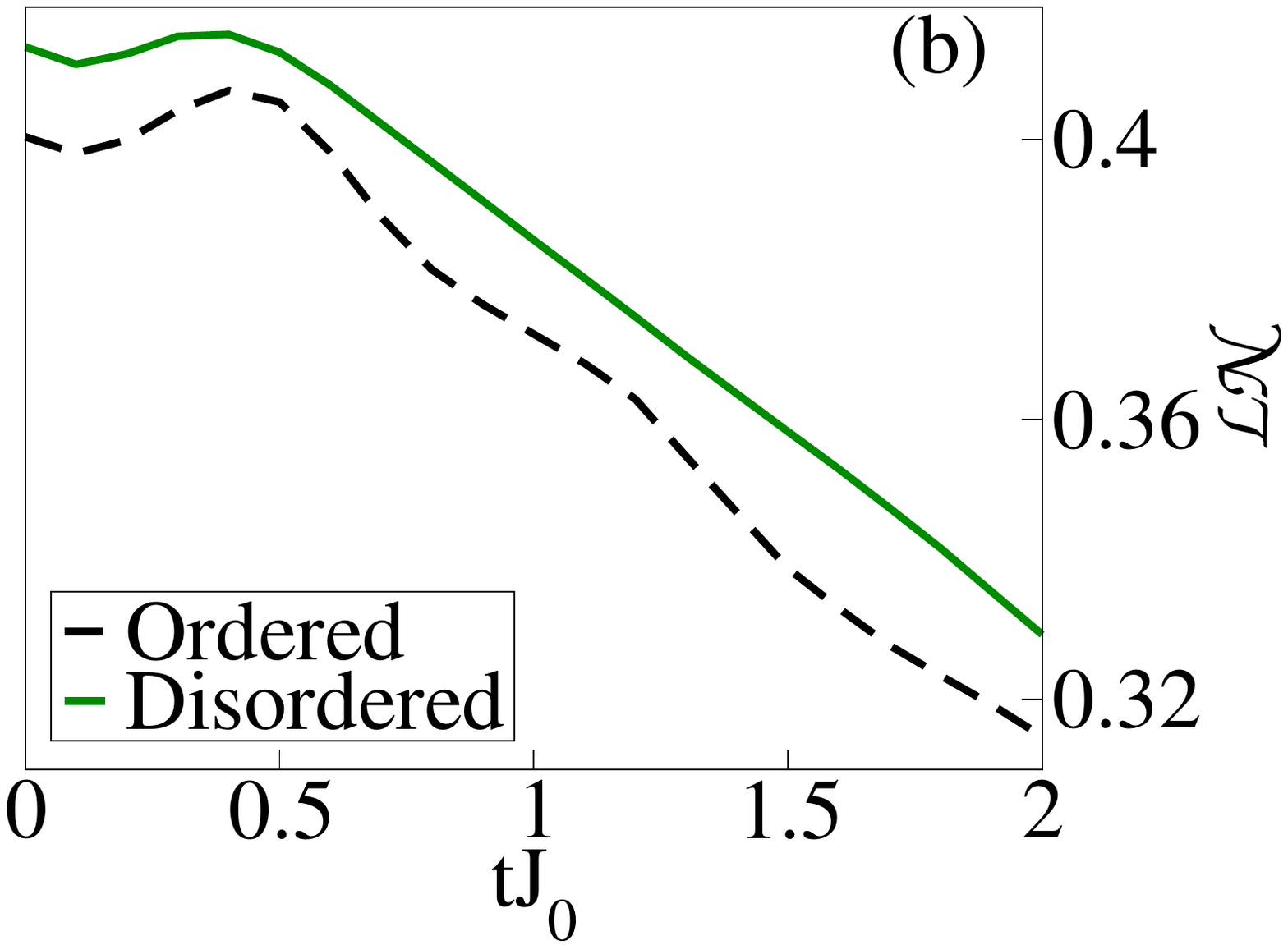}\\
    \includegraphics[width=0.5\linewidth]{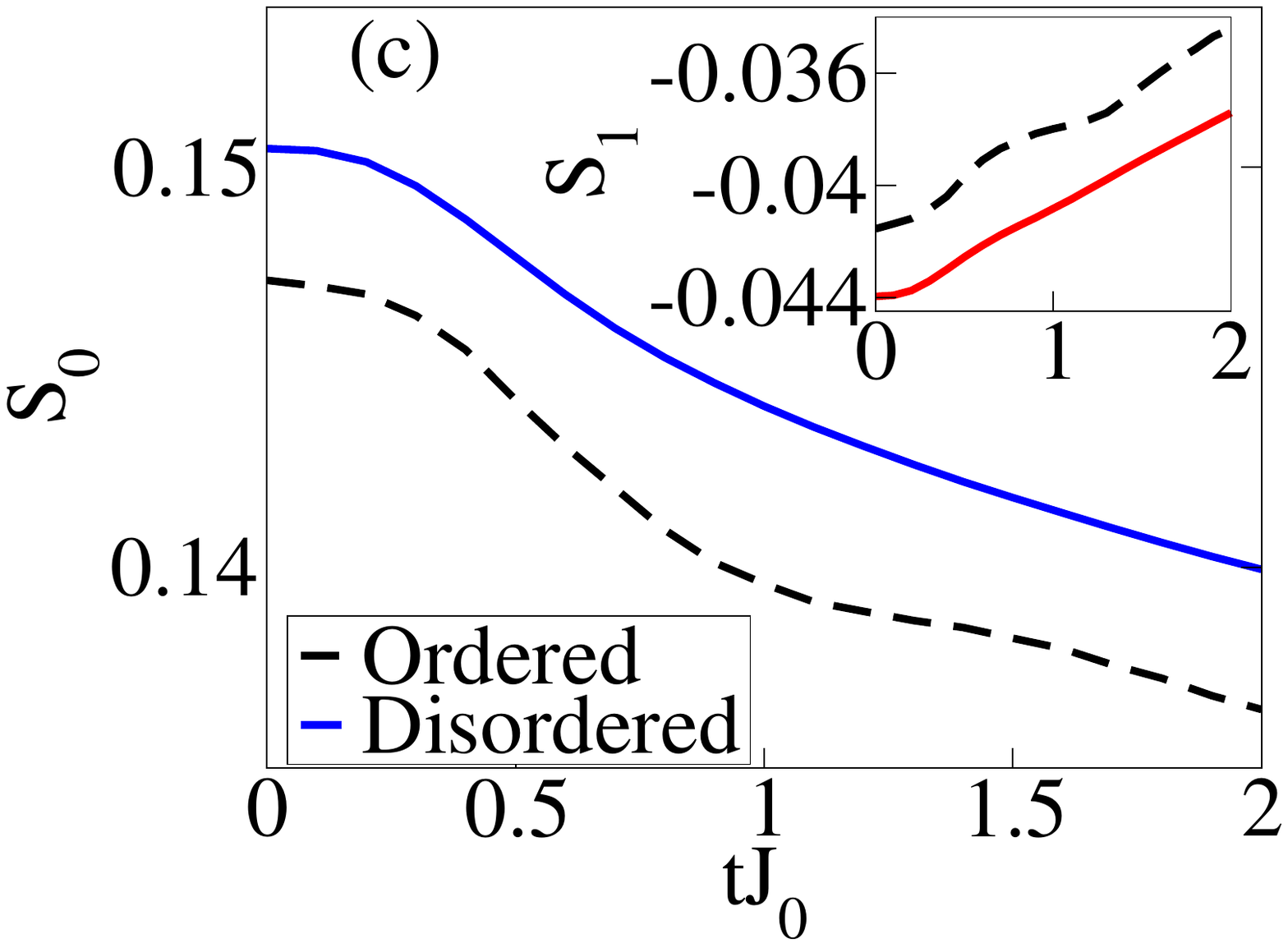}&
    \includegraphics[width=0.5\linewidth]{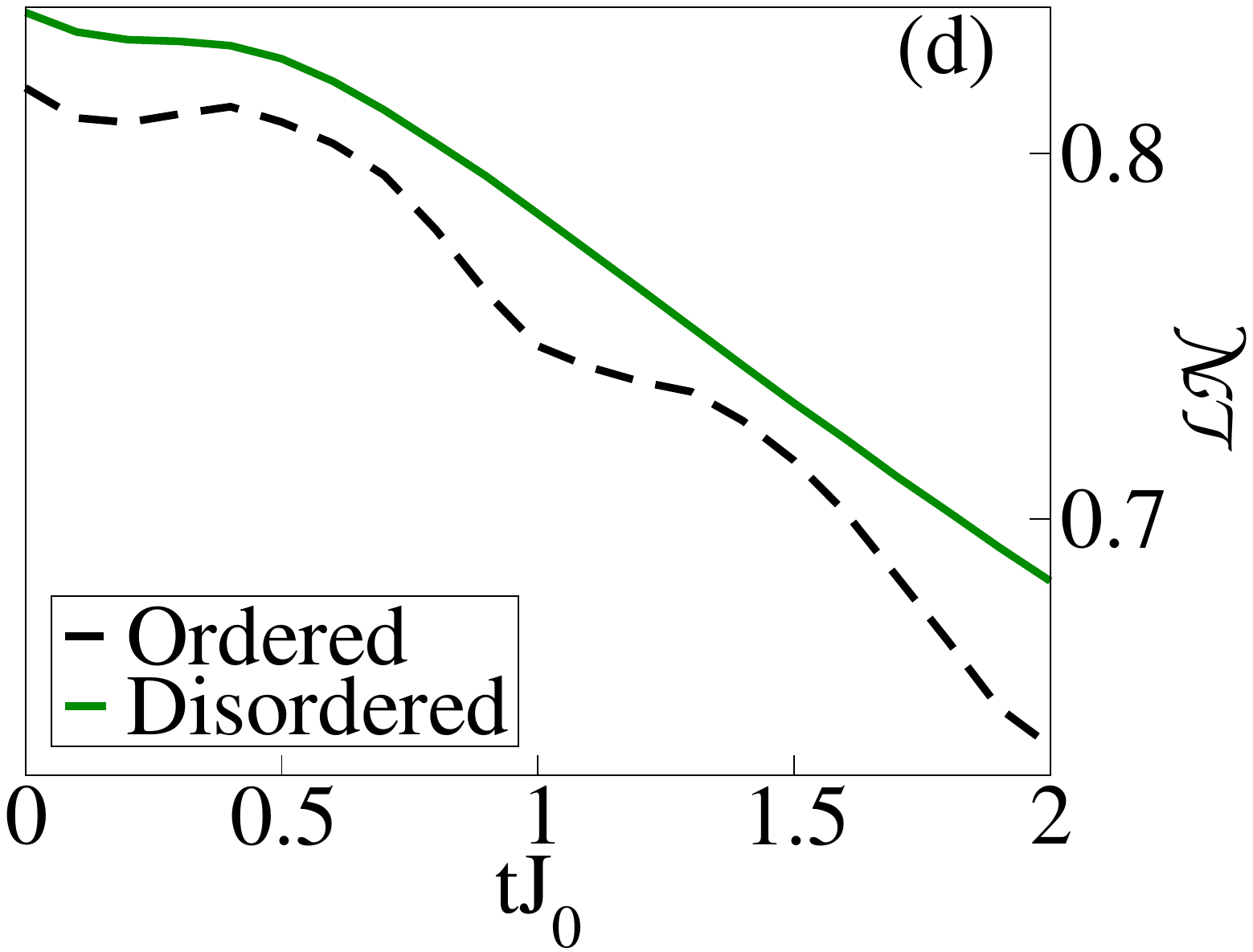}
  \end{tabular}
\caption{(a) Time evolution for the Bose-Hubbard Hamiltonian with an $N=8$ system showing ordered and disorder-averaged values of $M_1$ as a function of time for $\gamma=0.1J_0$, $2000$ quantum trajectories for the ordered state, and $50$ trajectories for each of $2000$ disorder realisations. The inset shows similar plot for $M_2$. (b) Time evolution for the Bose-Hubbard system showing ordered and disorder-averaged values of $\mathcal{LN}$ as a function of time for $\gamma=0.1J_0$ and $2000$ quantum trajectories for the ordered state and $50$ trajectories for each of $2000$ disorder realisations. (c) Time evolution for the Fermi-Hubbard Hamiltonian with an $N=8$ system showing ordered and disorder-averaged values of $S_0$ as a function of time for $\gamma=0.1J_0$ and $2000$ quantum trajectories for the ordered state and $50$ trajectories for each of $1500$ disorder realisations. The inset shows similar plot for $S_1$. (d) Time evolution for the same system showing ordered and disorder-averaged values of $\mathcal{LN}$ as a function of time for $\gamma=0.1J_0$ and $2000$ quantum trajectories for the ordered state and $50$ trajectories for each of $1500$ disorder realisations.}
\label{time}
\end{figure}

In the open system dynamics of the Bose-Hubbard model without disorder, the localisation due to spontaneous emission events is likely to increase the Mott orders as a function of time. This is what we observe in our calculation as is shown in figure~\ref{time}(a) where $M_1$ is plotted as the black dashed line with $U/J_0=2$ and $\gamma=0.1J_0$. The number of trajectories used for it is $2000$, making the error bars very small. The quenched disordered average is plotted as the solid blue line which is computed with disorder averaging with $2000$ realisations with $50$ trajectories for each value. The disorder-induced enhancement does not change much over time. Qualitatively, same features are found for $M_2$ which is shown in the inset. Figure~\ref{time}(b) depicts the same result for $\mathcal{LN}$ which is found to be decreasing over time in a similar fashion for both the ordered case and the quenched disordered average case, with the former displaying an undulating behaviour.

On the other hand, for the fermionic case, the decoherence causes the spin correlation function to diminish as a function of time. The value of $\Delta (S_0)$, however, is also found to be not changing much as a function of time as can be seen in figure~\ref{time}(c), which is computed on a $8$-site lattice with $2000$ trajectories for the ordered case with $U/J_0=2$, $\gamma=0.1J_0$ and $50$ trajectories for each of $1500$ disorder realisations in $J$. Again we omit the error bars as they are quite small. The inset of figure~\ref{time}(c) shows $\Delta (S_1)$ evolving with time which has similar nature to that of $\Delta (S_0)$. In the ordered case, $\mathcal{LN}$ for the two adjacent sites in the middle of the chain is found to be decreasing in time in an oscillatory fashion whereas the decrease for the value of the quenched disordered average has a steadier nature and it is always larger than the ordered value as a function of time (see figure~\ref{time}(d)). 

The strength of $\gamma$ determines how fast the observables decay in time. As long as $\gamma$ is small enough for the Born-Markov approximation to remain valid, we observe that both ordered and disorder-averaged values of the observables decay faster with increasing $\gamma$. However, the distance between those values, which quantifies the disorder-induced enhancement, does not show appreciable change as a function of $\gamma$. Therefore, we fix its value to $\gamma = 0.1 J_0$ while showcasing the dynamical behaviour.

% =============================================================================
\section{Site-dependent disorder}
\label{Site-dependent disorder}

\begin{figure}[t]
\centering
  \begin{tabular}{cc}
    \includegraphics[width=0.45\linewidth]{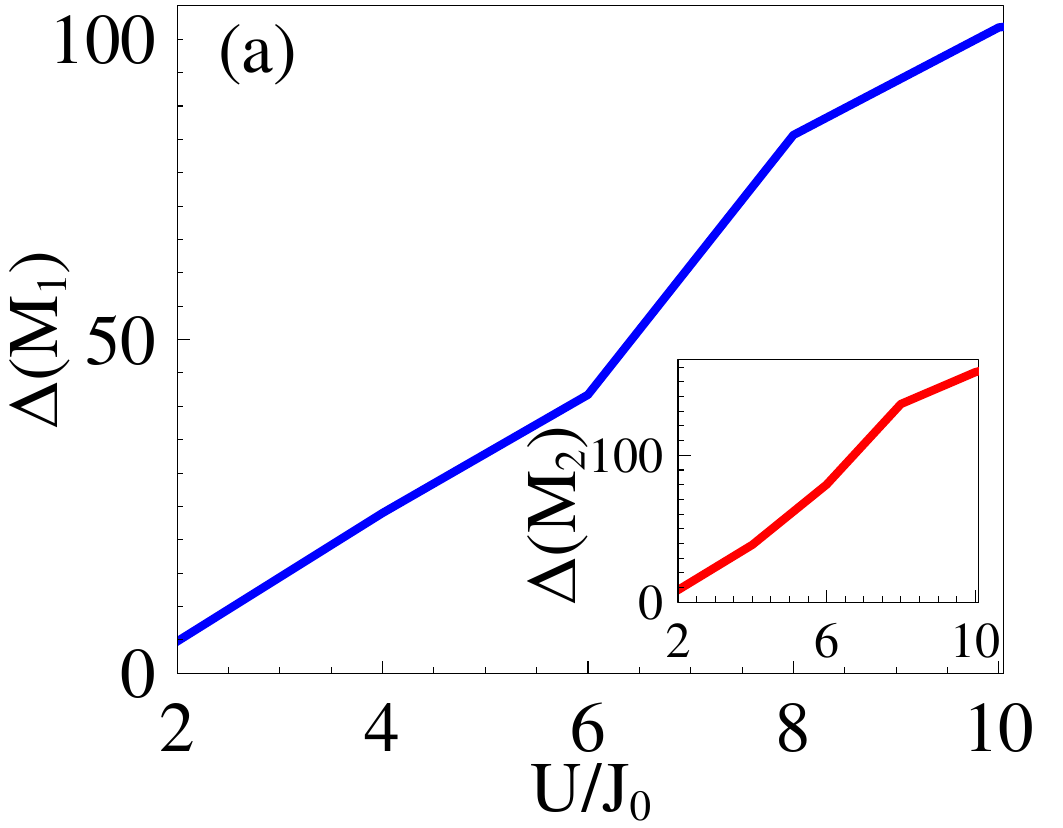}&
    \includegraphics[width=0.45\linewidth]{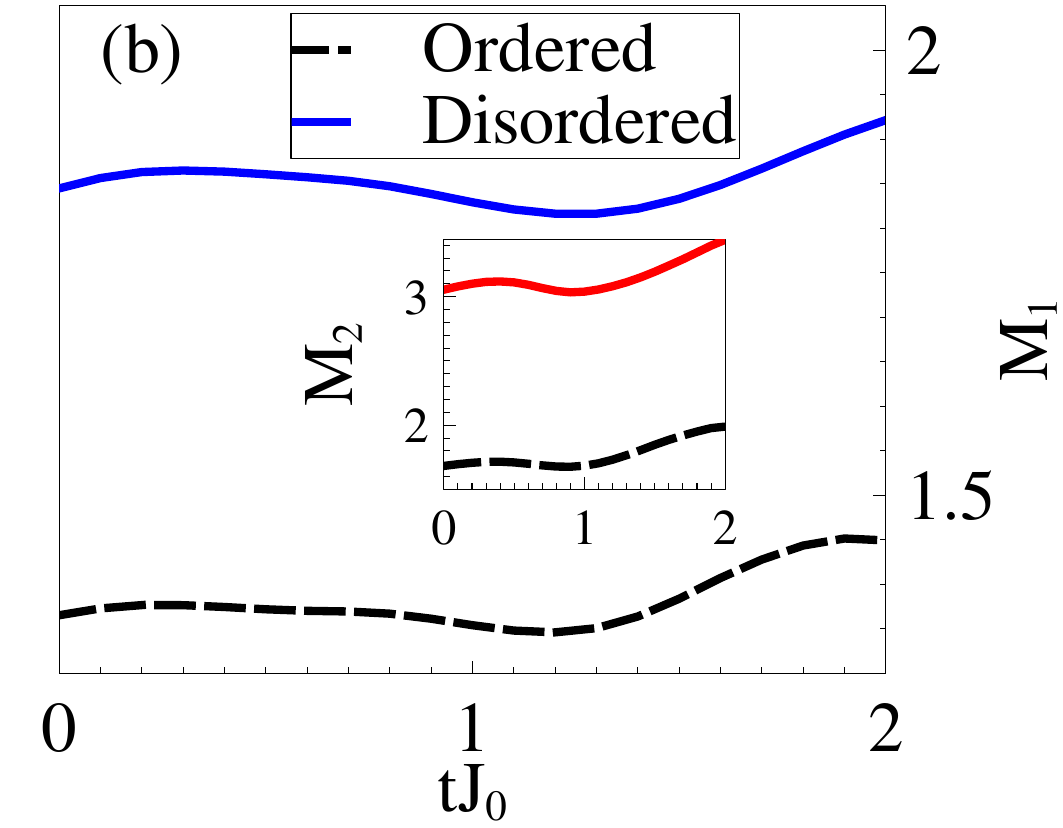}
  \end{tabular}
\caption{Effect of quenched disorder averaging in Bose-Hubbard model with site-dependent disorder in the tunneling terms. The system size is $N=8$, where we have used exact digonalisation with periodic boundary condition. (a) $\Delta (M_1)$ as a function of $U/J_0$, with $5000$ disorder realisations. The inset shows similar plot for $M_2$. (b) Time evolution of the ordered (dashed line) and disorder-averaged (solid line) value of $M_1$ for $U/J_0=5$, $\gamma=0.1J_0$, $50$ quantum trajectories for the ordered state, and $100$ trajectories for each of $5000$ disorder realisations. The inset shows similar plot for $M_2$.}
\label{site}
\end{figure}

We now present results for the case where we consider Hamiltonians in equations~\eqref{BH} and \eqref{FH}, with tunnelling parameters becoming site-dependent in the presence of the quenched disorder in each realisation. This can happen, for example, due to spatial fluctuations in the amplitude of the laser that generates the optical lattice. The tunnelling terms now take the form
\begin{align}
J_i = J_0 + \delta J_i(\mu,\sigma).
\end{align}
In this case, we find that the disorder-induced enhancement is absent for the spin-correlations in the Fermi-Hubbard model for all interaction strengths but is present for the Mott orders in the Bose-Hubbard model. This is shown in figure~\ref{site}(a) for a $8$-site system with unchanged $\mu$ and $\sigma$. In contrast to the earlier case, the enhancement in the Mott orders $M_1$ amd $M_2$ here actually increases with the interaction strength and is also significantly higher than before. We have obtained qualitatively similar results with DMRG calculations for a $32$-site system.  Next, we perform the open system dynamics using quantum trajectory method as before. We find that both the ordered values and the corresponding disordered averages decrease slightly initially and then start to increase with time. The disorder-induced enhancement stays almost unchanged. This is displayed in figure~\ref{site}(b) for $U/J_0=5$ and $\gamma=0.1J_0$. 

% =============================================================================
\section{Conclusion}
\label{Conclusion}

Remarkable level of control in ultracold atomic systems have led to the possibility of engineering different types of disorder and to simulate the resulting effects. We consider the case in which the system has a quenched disordered Hamiltonian parameter, where the equilibration time of the parameter is many orders of magnitude higher than the relevant observation times for the physical quantities of interest. The relevant physical quantities are then the quenched averaged ones. We computed the response of quenched disorder averaging in physical observables for Bose- and Fermi-Hubbard systems, where Gaussian random disorder is either present or inserted in the Hamiltonian parameters. For bosons, the observables that we considered are the single-particle density matrix elements that characterise the many-body ground state as well as the two-site nearest neighbour logarithmic negativity as a measure of bipartite entanglement. For fermions, we evaluated the spin correlation functions as measures of magnetic order along with logarithmic negativity. In both Bose- and Fermi-Hubbard systems, we observed disorder-induced enhancement of the observables in the ground state for weak interactions, when the disorder is introduced in the tunnelling term. Specifically, we observe that the values of such observables, averaged over the disorder realizations, is larger than their values in the case without disorder. We found that such features terminate when the on-site interaction becomes stronger. For the case of site-dependent disorder in tunnelling parameters, we found that the enhancement only occurs for bosons. 

The disorder-induced enhancements can be explained by analyzing the statistics of the disorder realisations and the dependence of the observables on the Hamiltonian parameters. We began by looking at the behaviour of the ingredients in the functional form of the physical observables as functions of the ratio of onsite interaction strength to the tunnelling rate, in the case when disorder is \emph{absent}. For low values of the ratio, within the superfluid regime, we observed that a quantity that is in the numerator of one of the physical observables of interest falls faster than in the cases of large values of the ratio. We then looked at the distribution of the ratio itself in the \emph{disordered} case, and found that it is also an asymmetric one. By analyzing the tilts in the contributions to the quenched average of the physical quantity due to the asymmetries in the above distributions, it was possible to pin down the reason for the disorder-induced enhancement of the physical quantity in the superfluid phase. A similar argument led to an understanding of the absence of such enhancement in the Mott phase for the same observable. The same line of analysis followed in all other cases of presence or absence of enhancement due to inclusion of disorder in tunnelling. The absence of enhancement specifically implies that the disorder-averaged values of the relevant observables are suppressed from their ordered values which are observed when there is no disorder.

We then considered the dynamics under the decohering action of spontaneous emission events. We found that the effects of the quenched disorder averaging do not change noticeably with respect to what we observe at the initial instance. 

Our results provide with a means to quantify the effects of such quenched disorder averaging and can be used to benchmark optical lattice experiments probing into the effects of disorder.  

% =============================================================================
\ack{We thank Aditi Sen(De) for useful discussions and Johannes Schachenmayer and Anton Buyskikh for helping with numerics. We acknowledge partial support from the Department of Science and Technology, Government of India, through the QuEST grant (grant number DST/ICPS/QUST/Theme-3/2019/120). SS acknowledges support from the Alberta Major Innovation Fund. Computational works were performed using the High Performance Scientific Computing facility at Harish-Chandra Research Institute, India.}

% =============================================================================
\section*{References}
\bibliographystyle{iopart-num}  %unsrt
\bibliography{Disorder}

% =============================================================================
\end{document}